\documentclass{article} 
\usepackage{nips13submit_e,times}
\pdfoutput=1
\usepackage{hyperref}
\usepackage{url}
\usepackage{graphicx, amsmath}
\usepackage[psamsfonts]{amssymb}
\usepackage{subfigure}
\usepackage{array}

\title{Third-Order Edge Statistics: Contour Continuation, Curvature,
  and Cortical Connections}

\author{
Matthew Lawlor\\
Applied Mathematics\\
Yale University\\
New Haven, CT 06520 \\
\texttt{matthew.lawlor@yale.edu} \\
\And
Steven W. Zucker \\
Computer Science \\
Yale University \\
New Haven, CT 06520 \\
\texttt{zucker@cs.yale.edu} \\
}

\nipsfinalcopy 

\begin{document}

\maketitle

\begin{abstract}
Association field models have attempted to explain human contour grouping performance, and 
to explain the mean frequency of long-range horizontal connections across cortical columns in V1.  
However, association fields only depend on the pairwise statistics of
edges in natural scenes.  We develop a spectral test of the
sufficiency of pairwise statistics and show there is significant
higher order structure.  An analysis using a probabilistic spectral
embedding reveals curvature-dependent components.

\end{abstract}


\section{Introduction}


Natural scene statistics have been used to explain a variety of neural structures.
Driven by the hypothesis that early layers of visual
processing seek an efficient representation of natural
scene structure,  decorrelating
or reducing statistical dependencies between subunits provides insight into 
retinal ganglion cells \cite{van1992}, cortical simple cells \cite{olshausen1996,bell1997}, 
and the firing patterns of larger ensembles \cite{vinje2000}. 
In contrast to these statistical models,
the role of neural circuits can be characterized functionally
\cite{BenShahar2004,Sato2012} by positing roles such as denoising,
structure  enhancement, and geometric computations.  Such models are based 
on evidence of excitatory connections among co-linear and co-circular neurons
\cite{Fitzpatrick2003}, as well as the presence of co-linearity and 
co-circularity of edges in natural images \cite{Geisler2001}, \cite{Elder1998}.  The fact that 
statistical relationships have a geometric structure is not surprising:  To the extent that 
the natural world consists largely of piecewise smooth objects, the boundaries of those objects should 
consist of piecewise smooth curves.  

Common patterns between excitatory neural connections, co-occurrence statistics, and 
the geometry of smooth surfaces suggests that the functional and statistical approaches can be 
linked.  Statistical questions about edge distributions in natural images have differential 
geometric analogues, such as the distribution of intrinsic derivatives in natural objects.
From this perspective, previous studies of natural image statistics have primarily 
examined  ``second-order'' differential properties of curves; i.e., the average change 
in orientation along curve segments in natural scenes.  The pairwise statistics 
suggest that curves tend toward co-linearity, in that the (average) change in
orientation is small.  Similarly, for long-range horizontal connections, cells with similar 
orientation preference tend to be connected to each other.

Is this all there is? From a geometric perspective, do
curves in natural scenes exhibit continuity in curvatures, or just in
orientation?  Are edge statistics well characterized 
at second-order? Does the same hold for textures? 

To answer these questions one needs to examine higher-order statistics
of natural scenes, but this is extremely difficult
computationally. One possibility is to design specialized patterns,
such as intensity textures \cite{tkavcik2010}, but it is difficult to
generalize such results into visual cortex. We make use of natural invariances in image
statistics to develop a novel spectral technique based on preserving a
probabilistic distance. This distance characterizes what is beyond
association field models (discussed next) to reveal the ``third-oder''
structure in edge distributions. It has different
implications for contours and textures and, more generally, for learning.




\section{Edge Co-occurrence Statistics}

Edge co-occurrence probabilities are well studied
\cite{august2000,Geisler2001,Elder2002,kruger1998}. Following them,
we use random variables indicating
edges at given locations and orientations.  More precisely, an edge at
position, orientation $(x_i,
y_i, \theta_i)$, denoted $E_i$, is a $\{0,1\}$ valued random variable.   
Co-occurrence statistics examine various aspects of pairwise marginal distributions,
which we denote by $P(E_i, E_j)$.  

The image formation process endows scene statistics with a natural translation 
invariance.  If the camera were allowed to rotate randomly about the focal axis, natural scene statistics would also have a rotational invariance.  For computational convenience, we 
enforce this rotational invariance by randomly rotating our images.  Thus, 
\[P(E_{(x_i, y_i, \theta_i)}, ... , E_{(x_j, y_j, \theta_j)}) = 
P(E_{(x_i + \delta x, y_i + \delta y, \theta_i + \delta \theta)}, ... , 
E_{(x_j + \delta x, y_j +\delta y, \theta_j + \delta \theta)}) \]

We can then estimate joint distributions of nearby edges by looking at \emph{patches} of edges
centered at a (position, orientation) location $(x_i, y_i, \theta_i)$ and rotating the patch into a canonical orientation
and position that we denote $E_0 = E_{(x_0, y_0, \theta_0)}$.

\[P(E_{(x_i, y_i, \theta_i)}, ... , E_{(x_j, y_j, \theta_j)}) = 
P(E_{(x_i - x_j, y_i - y_j, \theta_i - \theta_j)}, ... , 
E_{(x_0, y_0 , \theta_0)}) \]

\begin{figure}[h]
  \begin{center}
\begin{tabular}{ccc}
    \includegraphics[width=.2\textwidth]{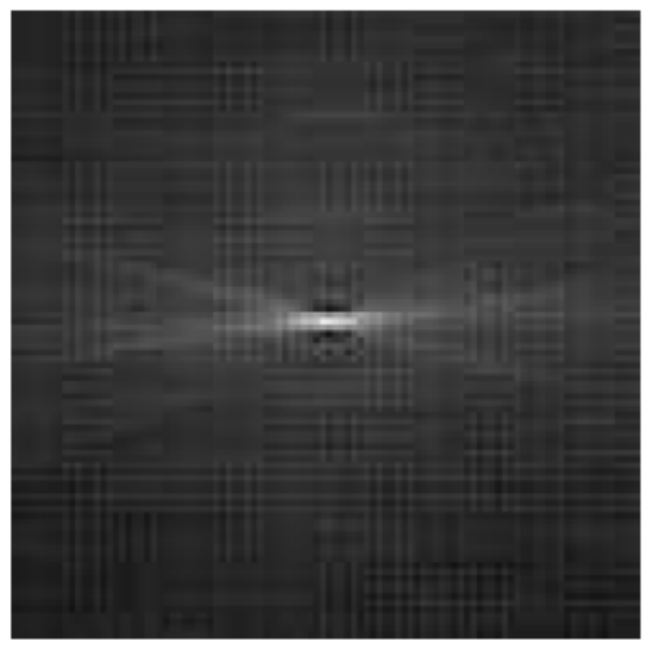} &
    \includegraphics[width=.2\textwidth]{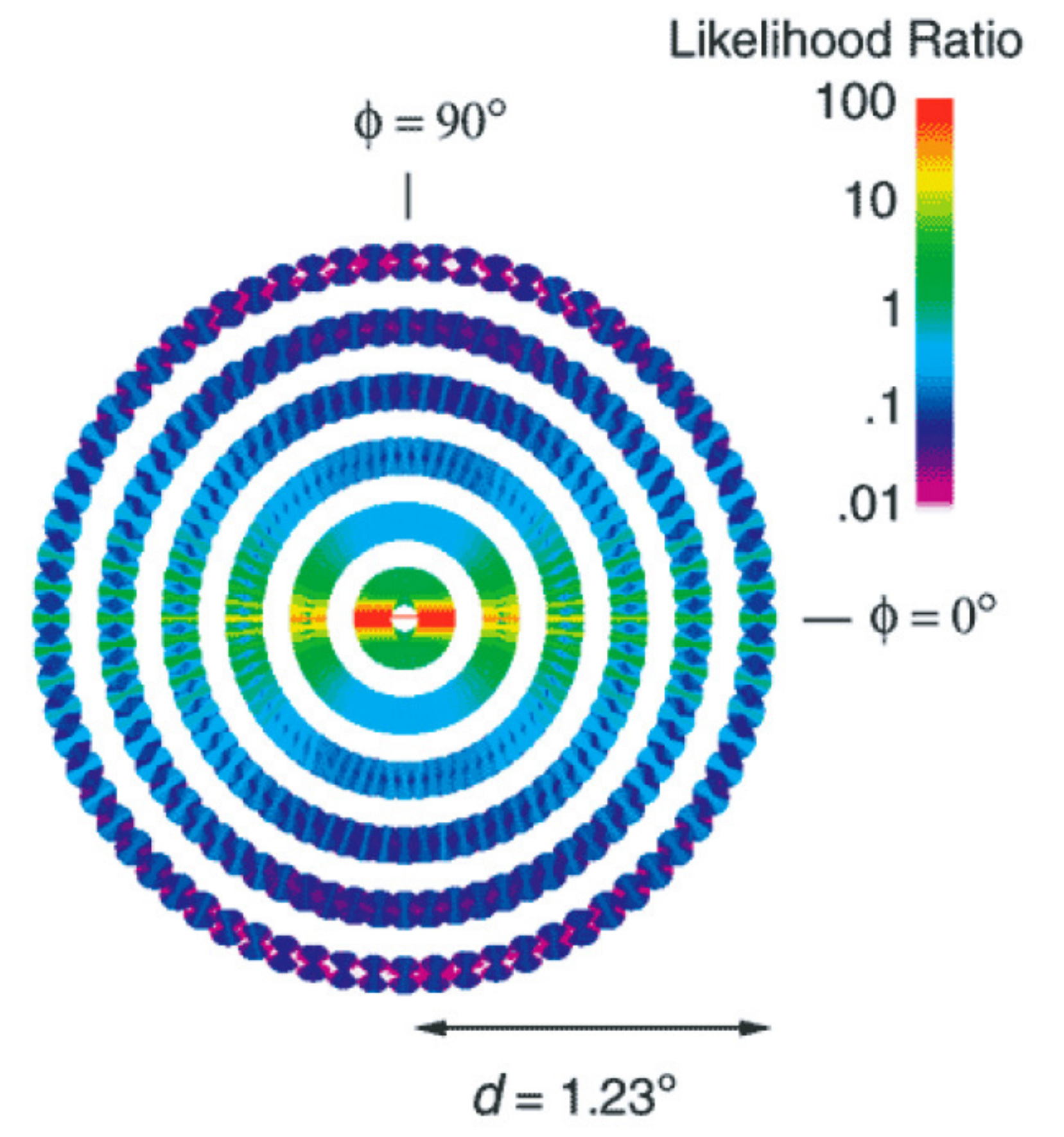} &
    \includegraphics[width=.2\textwidth]{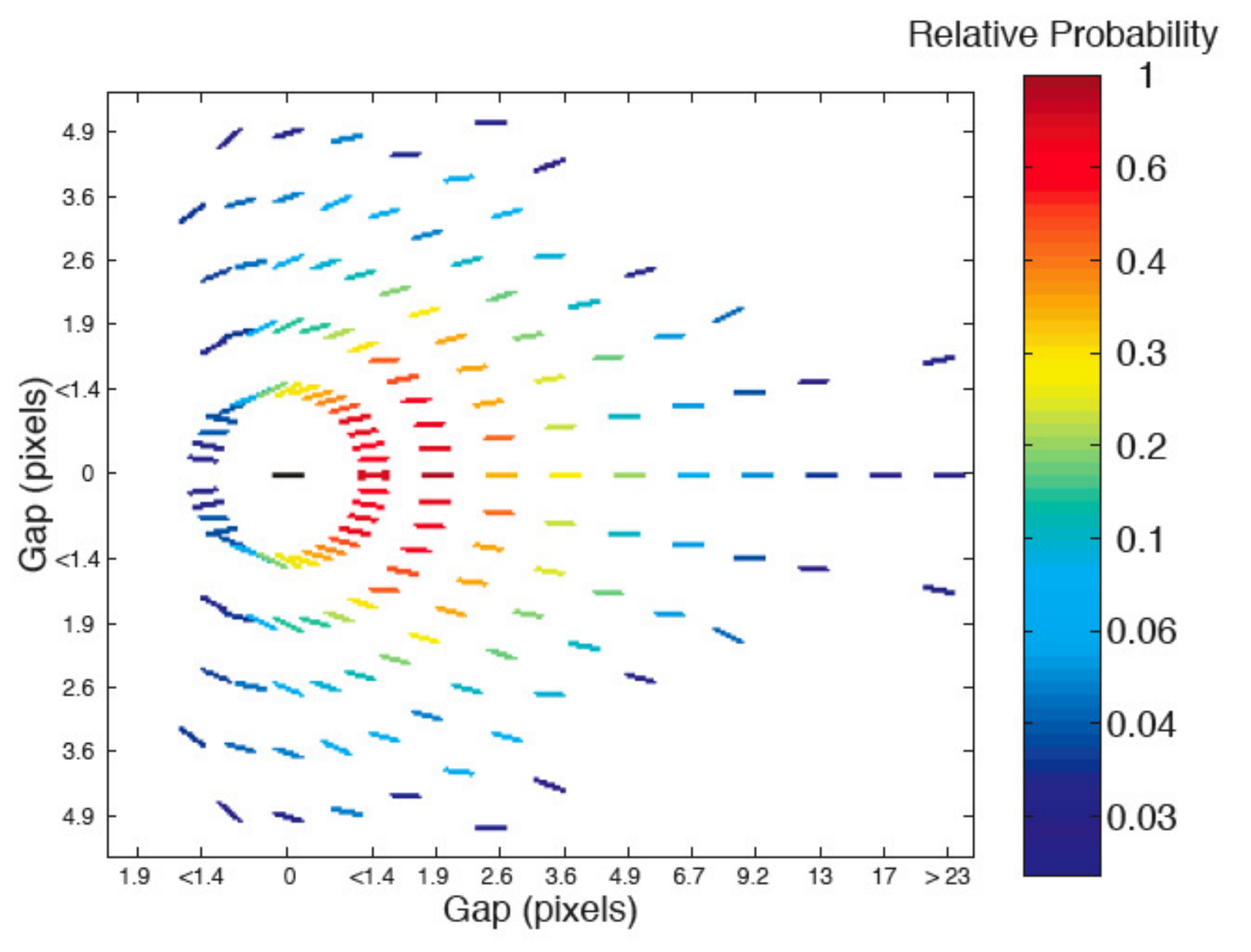} \\
August and Zucker, 2000  & Geisler et al, 2001 & Elder \& Goldberg, 2002 \\
\end{tabular}
 \end{center}
  \caption{Association fields derive from image co-occurrence
    statistics. We interpret them as illustrating the probability
    (likelihood) of an edge near a horizontal edge at the
    center position.}
  \label{fig:august-geisler-elder}
\end{figure}

Several examples of statistics derived from the distribution of $P(E_i, E_0)$ 
are shown in Fig.~\ref{fig:august-geisler-elder}.  
These are pairwise statistics of oriented edges in natural images.  
The most important visible feature of these pairwise statistics is that of \emph{good continuation}:  Conditioned on the presence of an edge at the center, edges of similar orientation and horizontally aligned with the edge at the center have high probability.  Note 
that all of the above implicitly or explicit enforced rotation invariance, either by only examining relative orientation with respect to a reference orientation or by explicit rotation of the images.  

\begin{figure}
\centering
\begin{tabular}{cc}
\includegraphics[width=.30\textwidth]{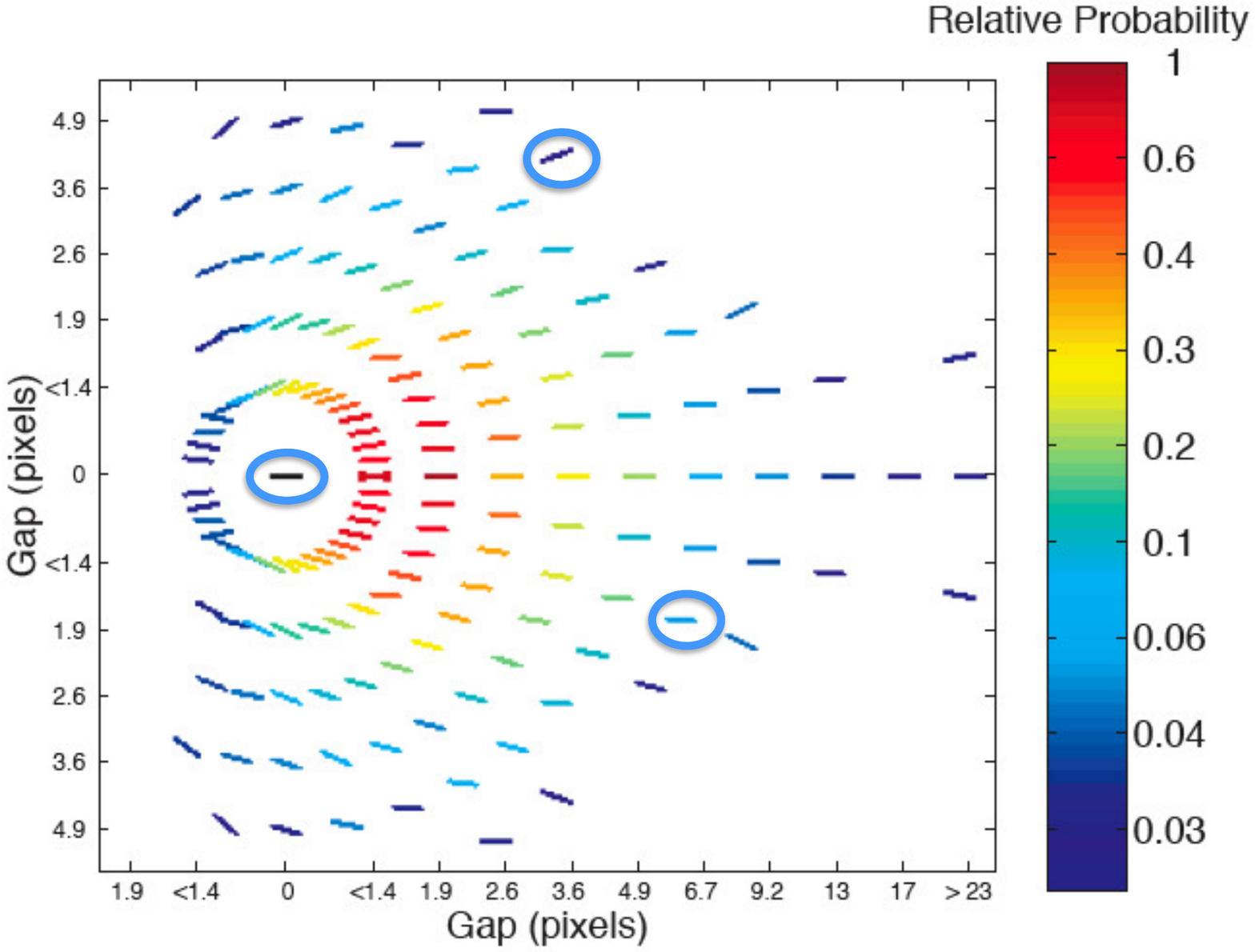} &
\includegraphics[width=.30\textwidth]{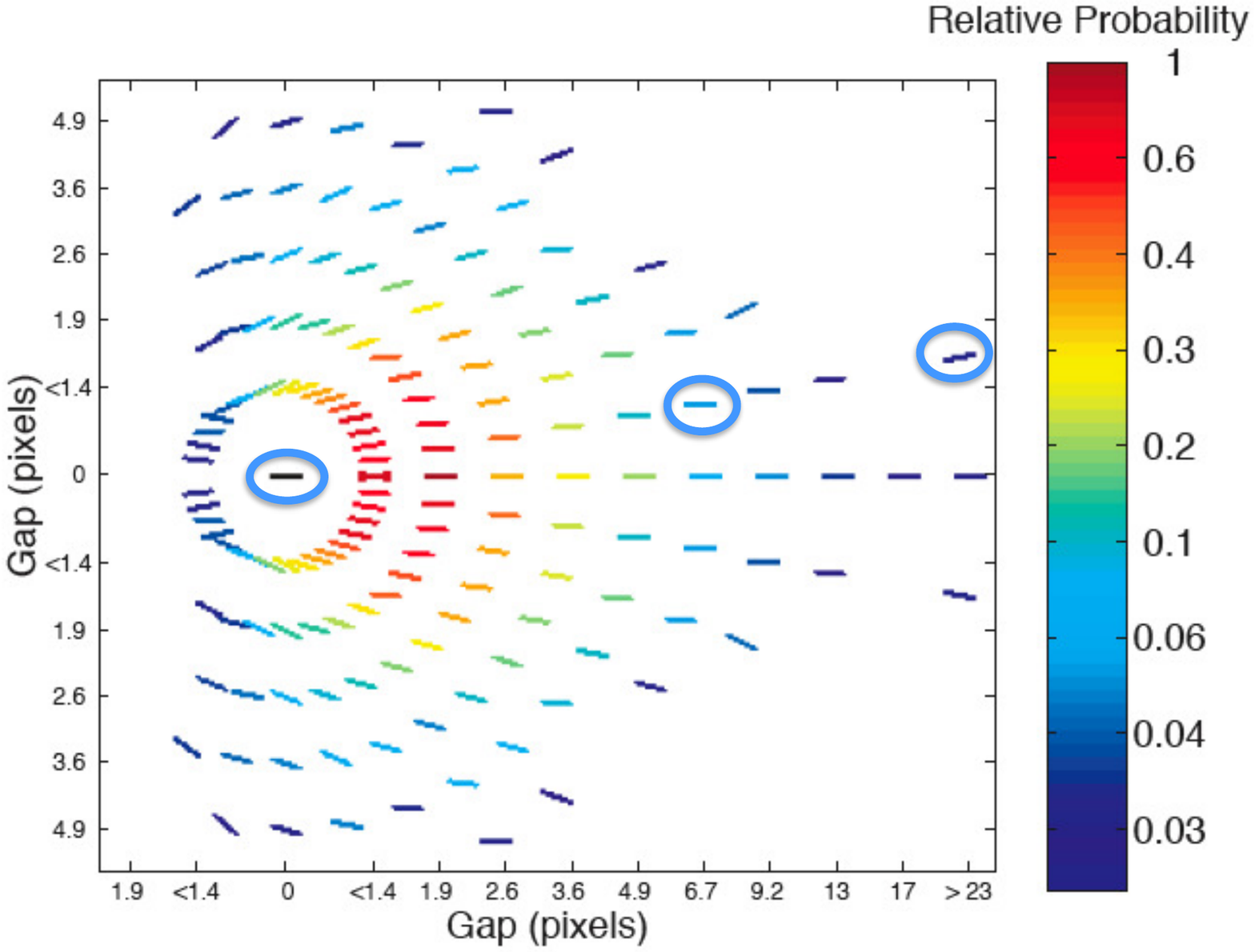}	
\end{tabular}			
\caption{Two approximately equally likely triples of edges under the pairwise independence assumption of
Elder et. al.  Conditional independence is one of several possible
pairwise distributional assumptions.  Intuitively, however, the second
triple is much more likely.  We examine third-order statistics to 
demonstrate that this is in fact the case.}
  \label{fig:elder-choices}
\end{figure}

It is critical to estimate the degree to which these pairwise statistics characterize the 
full joint distribution of edges (Fig.~\ref{fig:elder-choices}).  Many models for neural firing patterns imply relatively low
order joint statistics.  For example, spin-glass models imply pairwise statistics are sufficient, while Markov random fields have an order determined by the size of neighborhood cliques.

\section{Contingency Table Analysis}
To test whether the joint distribution of edges can be well described
by pairwise statistics, we performed a contingency table analysis of
edge triples at two different threshold levels.  We computed estimated
joint distributions for each  triple  of edges in an $11 \times 11
\times 8$ patch, not constructed to have an edge at the center.
Using a $\chi^2$ test, we computed the probability that each edge triple distribution could occur under hypothesis $H_0:\{\text{No three way interaction}\}$.  
This is a test of the hypothesis that 
\[\log P(E_i, E_j, E_k) = f(E_i, E_j) + g(E_j, E_k) + h(E_i, E_k)\] 
for each triple $(E_i, E_j, E_k)$, and includes the cases of independent edges, conditionally 
independent edges, and other pairwise interactions.
{\em For almost all triples, this probability was extremely small.}
(The few edge triples for which the null hypothesis cannot be rejected consisted of edges that 
were spaced very far apart, which are far more likely to be nearly statistically independent of one another.)

	\vspace{5ex}

	\begin{tabular}{l|ll}
		$n = 150705016$ & threshold $= .05$ & threshold $= .1$ \\
		\hline
		percentage of triples where $p_{H_0} >$ .05 & 0.0082\% & 0.0067\%
	\end{tabular}

\section{Counting Triple Probabilities}\label{triplesection}

We chose a random sampling of black and white images from the van Hataren image dataset\cite{hateren1998}.  They were randomly rotated and then filtered using oriented Gabor filters
covering 8 angles from $[0,\pi)$.  Each Gabor has a carrier period of
1.5 pixels per radian and an envelope standard deviation of 5 pixels.
The filters were convolved in near quadrature pairs, squared and summed.  To restrict
analysis to the statistics of curves, we applied local non-maxima suppression across 
orientation columns in a direction normal to the given orientation.
This removes effects of filter blurring
and oriented texture patterns (considered shortly). The resulting
edge maps were subsampled to eliminate statistical dependence due to overlapping filters. 

\vspace{1ex}
Thresholding the edge map yields
$E:U \rightarrow \lbrace 0,1 \rbrace$,  where $U \subset \mathbb{R}^2\times \mathbb{S}$ is a discretization of $\mathbb{R}^2\times \mathbb{S}$. We treat $E$
as a function or a binary vector as convenient.  We randomly select
$21\times 21 \times 8$ image patches with an oriented edge at the
center, and denote these characteristic patches by $V_i$

\begin{figure}[h]
  \begin{center}
\begin{tabular}{cc}
    \includegraphics[width=.25\textwidth]{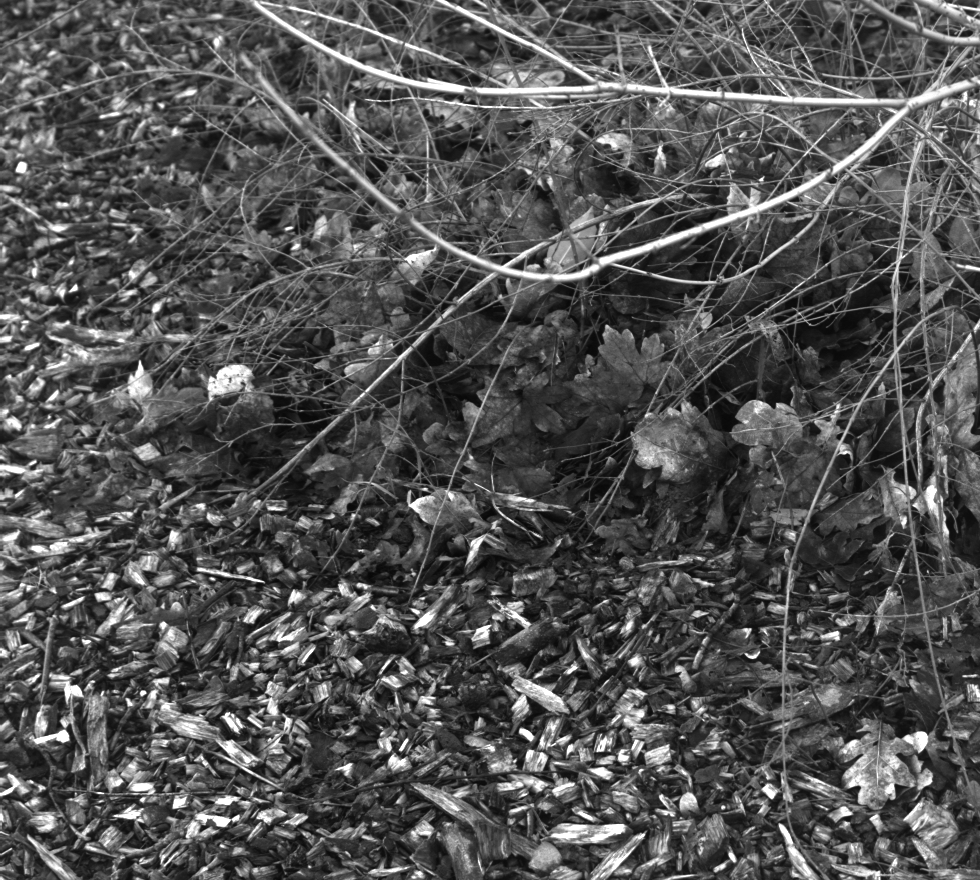} &
    \includegraphics[width=.25\textwidth]{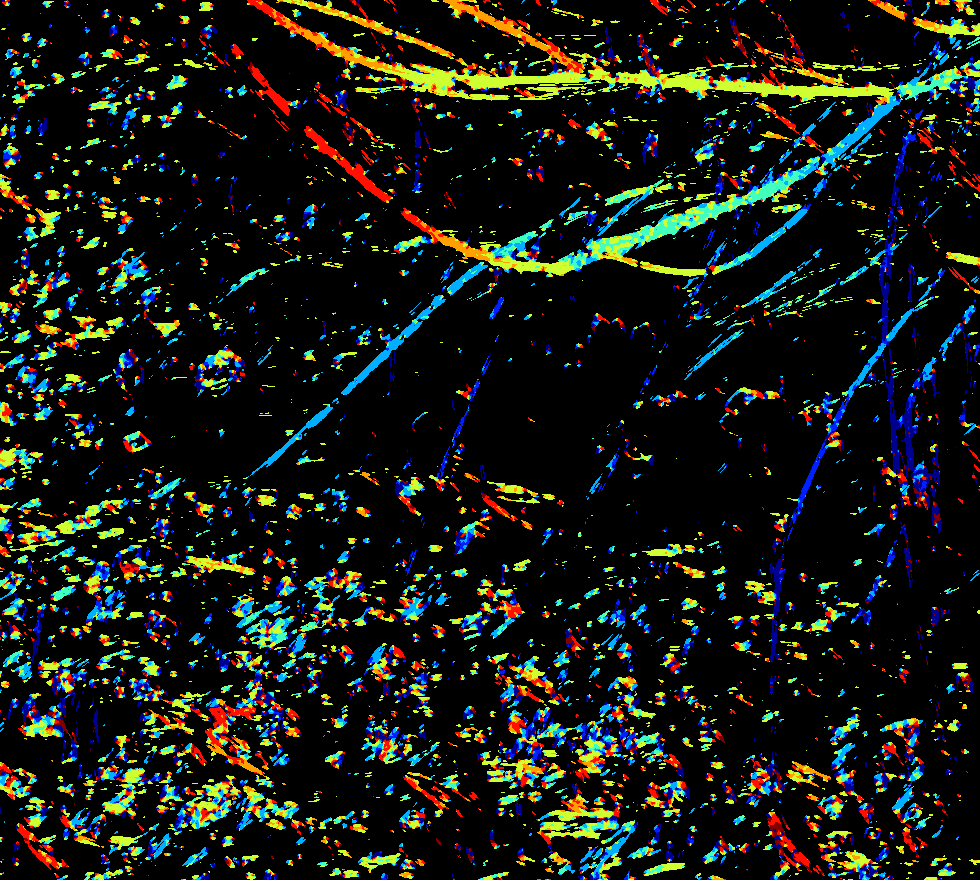} \\
 (a) & (b) \\
\end{tabular}
 \end{center}
  \caption{Example image (a) and edges (b) for statistical analysis. Note: color corresponds to orientation}
  \label{fig:edgesforstats}
\end{figure}

Since edges are significantly less frequent than their absence, we
focus on (positive) edge co-occurance statistics.  
For simplicity, we denote $P(E_i = 1, E_j = 1, E_k =1)$ by $P(i,j,k)$
(A small orientation anisotropy has been reported in
natural scenes (e.g., \cite{hansen2004}), but does not appear in our
data because we effectively averaged over orientations by randomly rotating the
images.)

We compute the matrix $P(i,j|0)$ where 

\begin{align*}
	P(i,j|0) &= P(E_i = 1, E_j = 1 | E_0 = 1)\\
	&\sim \frac{1}{n}\sum_{i = 1}^n V_iV_i^T
\end{align*}

where $V_i$ is a (vectorized) random patch of edges centered around an edge with orientation $\theta_i = 0$.

\section{Visualizing Triples of Edges}

By analogy with the pairwise analysis above, we seek to find those
edge triples that frequently co-occur. But this is significantly more
challenging.  For pairwise statistics, one simply fixes an edge to lie
in the center and ``colors'' the other edge by the joint probability of
the co-occurring pair (Fig.~\ref{fig:august-geisler-elder}). 
No such plot exists for triples of edges.  Even after conditioning, there are over 12 million 
edge triples to consider.  

Our trick: \emph{Embed edges in a low dimensional 
space such that the \emph{distance} between the edges represents the relative likelihood of co-occurrence.}
We shall do this in a manner such that
{\em distance in Embedded Space $\sim$ Relative Probability}.

As before, let $E_i$ be a binary random variable, where $E_i = 1$ means there
is an edge at location $v_i = (x_i, y_i, \theta_i)$.
We define a distance between edges 
\begin{align*}
D^2_+(v_i, v_j) = P(i,i|0) - 2P(i,j|0) + P(j,j|0)
\end{align*}

The first and the last terms represent pairwise
co-occurrence probabilities; i.e., these are the association
field. The middle term represents the interaction between $E_i$ and
$E_j$ conditioned on the presence of $E_0$.  
Thus this distance is zero if the edges always co-occur in images,
given the horizontal edge at the origin, and is large if the pair of
edges frequently occur with the horizontal edge but rarely together.
(The relevance to learning is discussed below.)

We will now show how, for natural images, edges can be placed in a low dimensional space where
the distance in that space will be proportional to this probabilistic distance.  

\section{Dimensionality Reduction via Spectral Theorem}

Exploiting the fact that $P(i,j|0)$ is a positive semi-definite matrix, we introduce the spectral expansion
\begin{equation*}
  P(i, j|0) = \sum_{l=1}^n \lambda_l \phi_l(i) \phi_l(j)
\end{equation*}
where $\phi_l$ is an eigenvector of the $K$ matrix.  

Define the
{\sc \large spectral embedding}
$\Phi:\left(
\begin{array}{c}
x_i \\
y_i \\
\theta_i 
\end{array} \right ) \rightarrow \mathbb{R}^n $

\begin{equation}
  \Phi(v_i) = \{ \sqrt{ \lambda_1 }\phi_1(i), \sqrt{ \lambda_2 }\phi_2(i), ... , \sqrt{ \lambda_n }\phi_n(i) \}
\end{equation}

The Euclidean distance between embedded points is then
\begin{align*}
\|\Phi(v_i) - \Phi(v_j) \|^2 & = \langle \Phi(v_i), \Phi(v_i) \rangle - 
2 \langle \Phi(v_i), \Phi(v_j) \rangle + \langle \Phi(v_j), \Phi(v_j)\rangle \\
& = P_+(i,i|0) - 2P_+(i,j|0) + P_+(j,j|0) \\
& = D^2_+(v_i,v_j)
\end{align*}

$\Phi$ maps edges to points in an embedded space where squared distance is equal to relative 
probability; we refer to this as diffusion distance.

\begin{figure}
	\begin{center}
	\includegraphics[width=.25\textwidth]{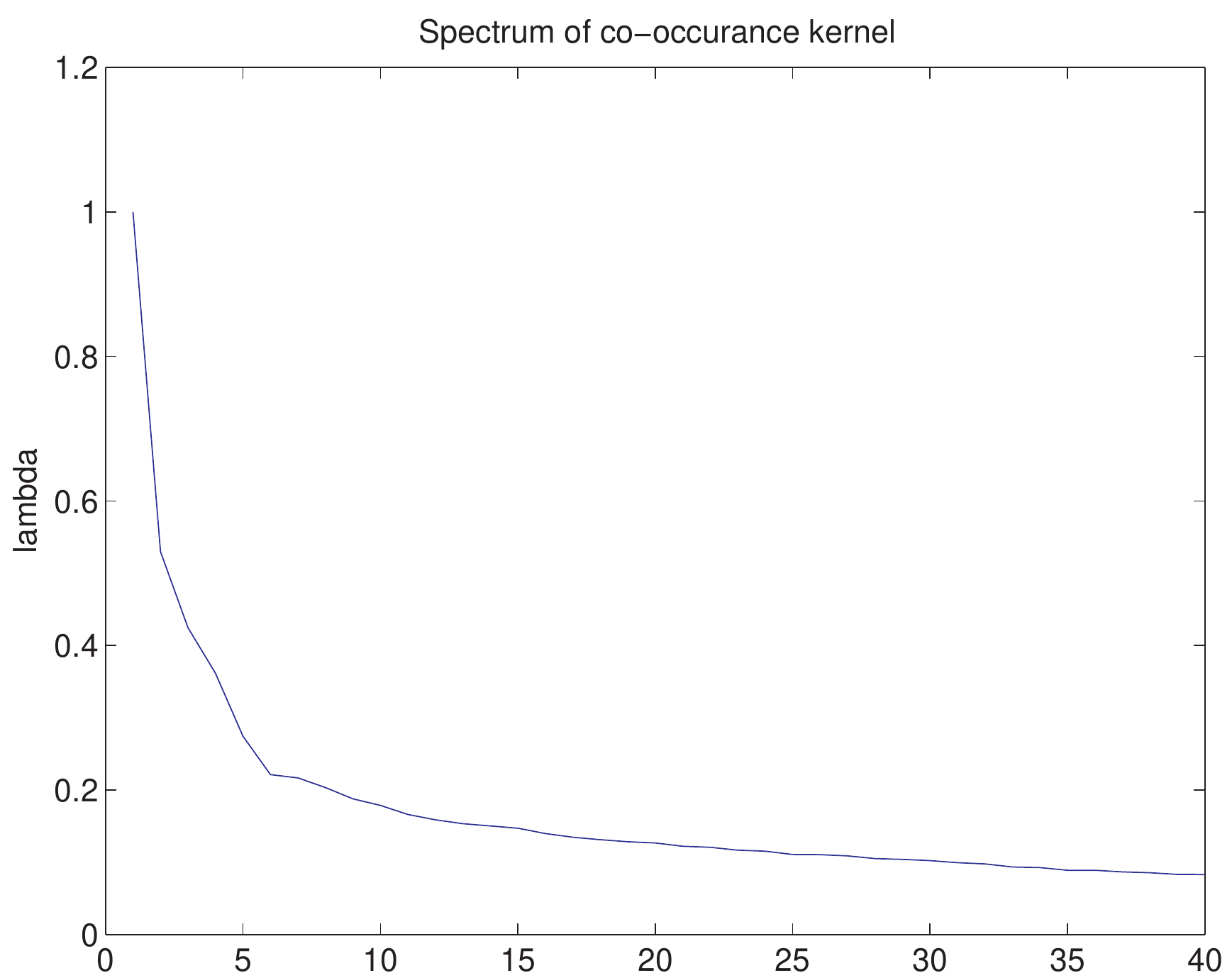}
	\caption{ Spectrum of $P(i,j|0)$.  Other spectra are similar.  Note rapid decay of the
	spectrum indicating the diffusion distance is well captured by
	embedding using only the first few eigenfunctions.}
	\end{center}
\label{fig:eigenvalues-edges}
\end{figure}

The usefulness of this embedding comes from the fact that the spectrum
of $P(i,j,0)$ decays rapidly (Fig.~\ref{fig:eigenvalues-edges}).  Therefore we truncate $\Phi$, including only dimensions with high eigenvalues.  This
gives a dramatic reduction in dimensionality, and allows us to visualize the relationship between triples of edges (Fig.~\ref{fig:eigenvalues-edges}).

\begin{figure}[h]
	\begin{tabular}{ccc}
         \includegraphics[width=.3\textwidth,clip=true, trim=0 1.5 0 1.5]{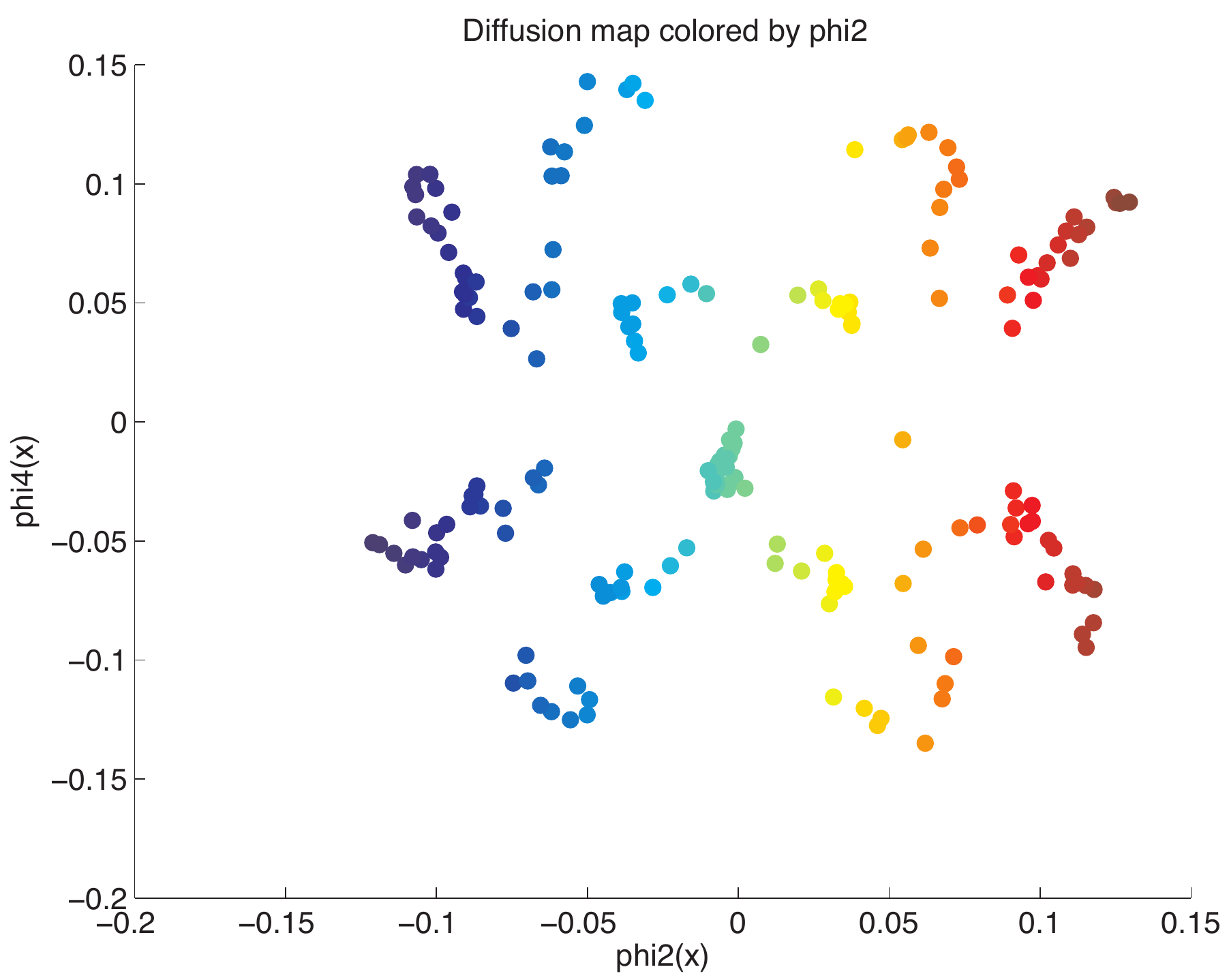} &
        \includegraphics[width=.3\textwidth,clip=true, trim=0 1.5 0 1.5]{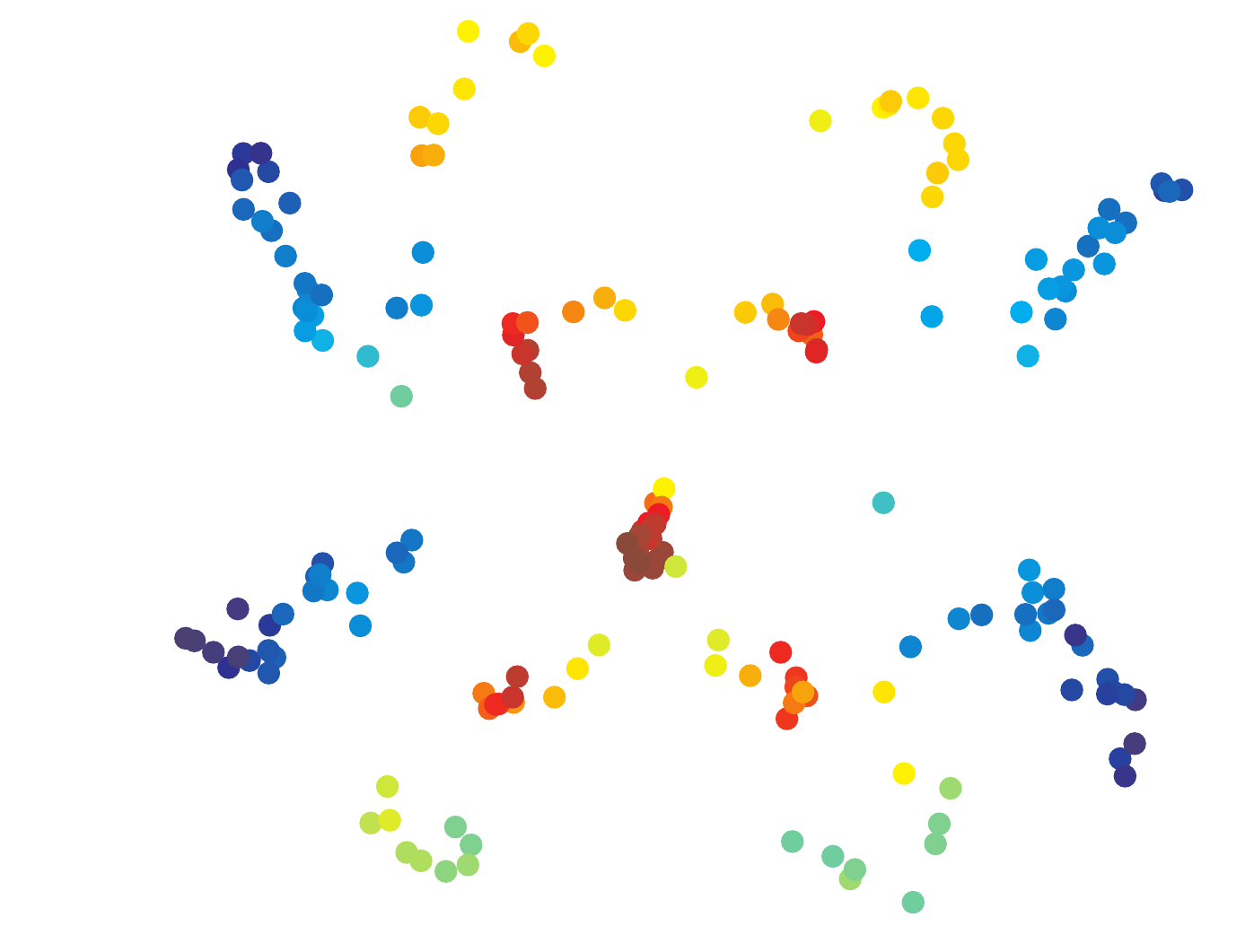} &
        \includegraphics[width=.3\textwidth,clip=true, trim=0 1.5 0 1.5]{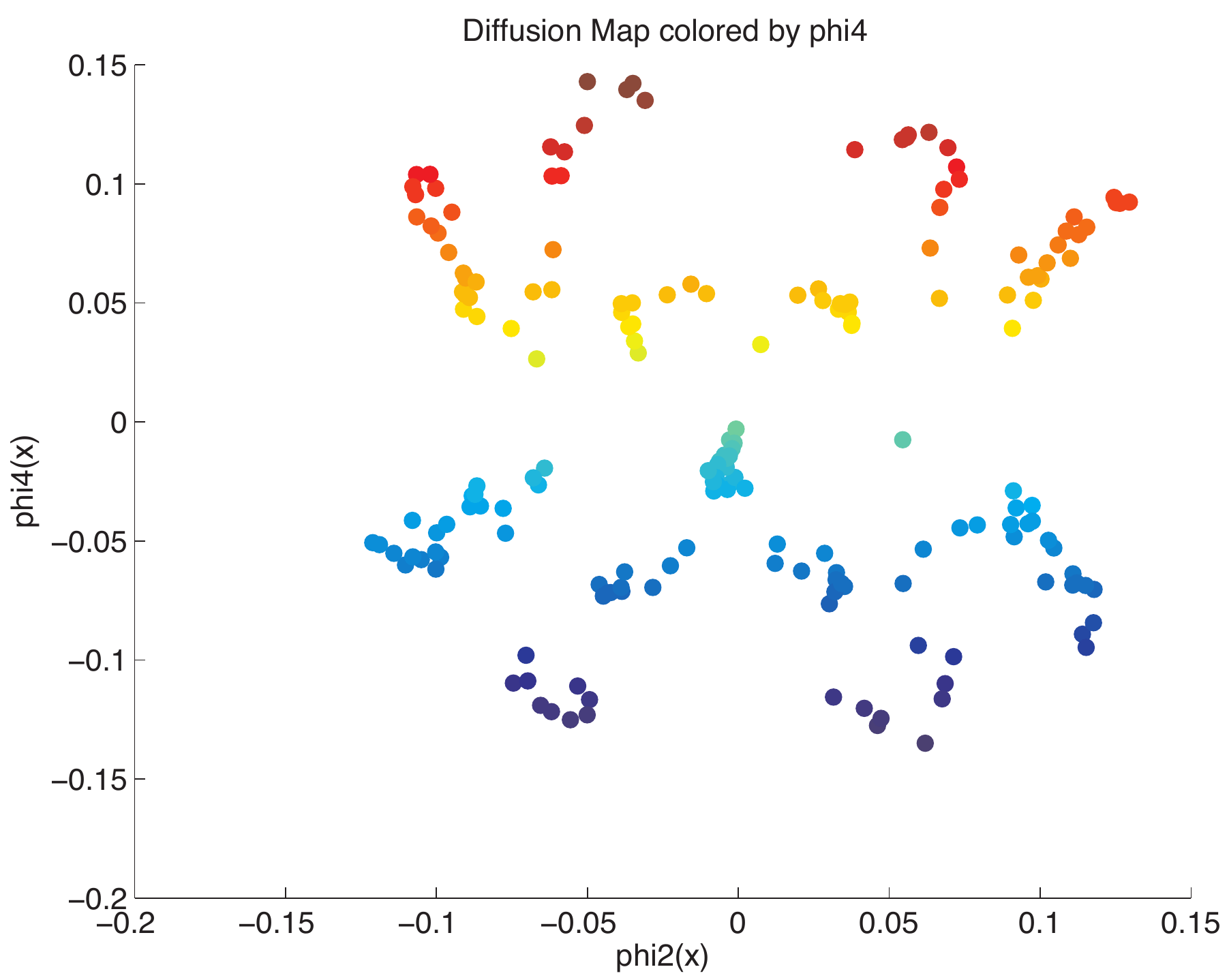}\\
        \includegraphics[width=.3\textwidth,clip=true, trim=0 1.5 0 1.5]{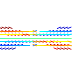} &
        \includegraphics[width=.3\textwidth,clip=true, trim=0 1.5 0 1.5]{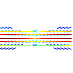} &
        \includegraphics[width=.3\textwidth,clip=true, trim=0 1.5 0 1.5]{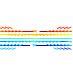} \\
        $\phi_2$ & $\phi_3$ & $\phi_4$
	\end{tabular}
	\caption{Display of third-order edge structure showing how
 oriented edges are related to their spectral embeddings. (top)
 Spectral embeddings. Note clusters of co-occuring edges. (bottom) Edge distributions. 
	The eigenvectors of $P(i,j,0)$ are used to color both 
	the edges and the embedding.  Clusters of edges in this embedding have a \emph{high probability of co-occurring} along with the edge in the center.  Compare with Fig.~\ref{fig:august-geisler-elder} where red edges all have high probability of occurring with the center, but no information is known about their co-occurrence probability.}
\label{fig:embeddings-1}
\end{figure}

To highlight information not contained in the association field, we normalized our
probability matrix by its row sums, and removed all low-probability edges.
Embedding the mapping from $\mathbb{R}^2 \times \mathbb{S} \rightarrow
\mathbb{R}^m$ reveals the cocircular structure of edge triples in the
image data (Fig.~\ref{fig:embeddings-1}.  
The colors along each column correspond, so similar colors map to
nearby points along the dimension corresponding to the row.  Under
this dimensionality reduction, each small cluster in diffusion space
corresponds to half of a cocircular field.  
In effect, the coloring by $\phi_2$ shows good continuation in
orientation (with our crude quantization) while the coloring by
$\phi_4$ shows co-circular connections. In effect, then, the
association field is the union of co-circular connections, which also
follows from marginalizing the third-order structure away.
 We used 40,000 
($21 \times 21 \times 8$) patches.  

\section{Implications: ``Cells that fire together wire together''}

Shown above are low dimensional projections of the diffusion map and their corresponding colorings in $\mathbb{R}^2 \times \mathbb{S}$.  
To provide a neural interpretation of these results, let
each point in $\mathbb{R}^2 \times \mathbb{S}$ represent a neuron with a receptive field centered at the point $(x,y)$ with preferred orientation $\theta$.  
Each cluster then signifies those neurons that have a high probability
of co-firing given that the central neuron fires, so clusters in
diffusion coordinates should be ``wired'' together by the Hebbian postulate.
Such curvature-based facilitation can explain the non-monotonic variance in excitatory long-range horizontal connections in V1~\cite{BenShahar2004,bosking1997}.
It may also have implications for the receptive fields of V2 neurons.  As clusters of co-circular V1 complex cells are correlated in their firing, it may be efficient to represent them 
with a single cell with excitatory feedforward connections. This
predicts that  efficient coding models that take high order interactions into account should exhibit cells tuned to 
curved boundaries.

\begin{figure}[h]
\centering
\begin{tabular}{cc}
\includegraphics[width=.3\textwidth]{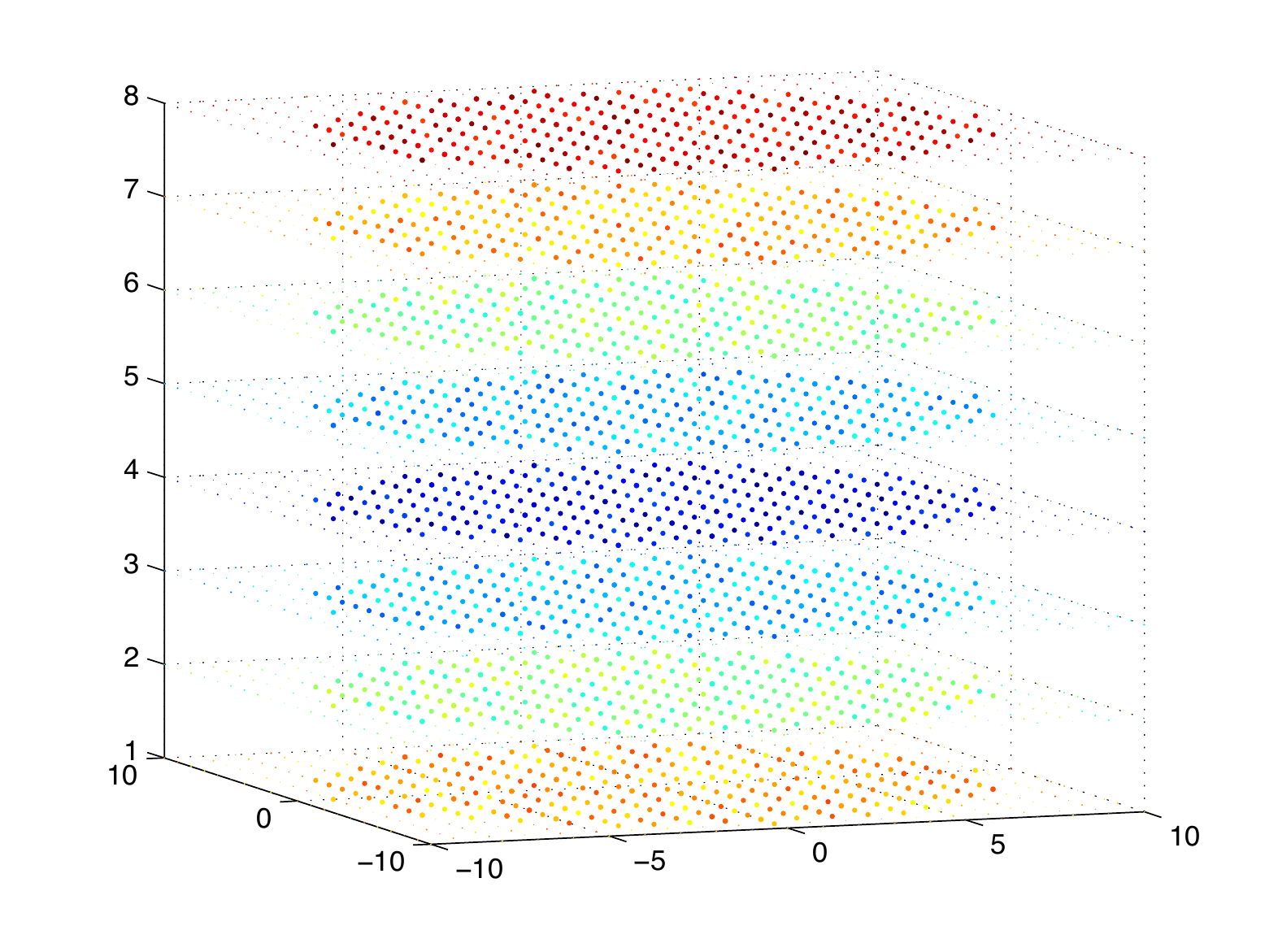} &
\includegraphics[width=.3\textwidth]{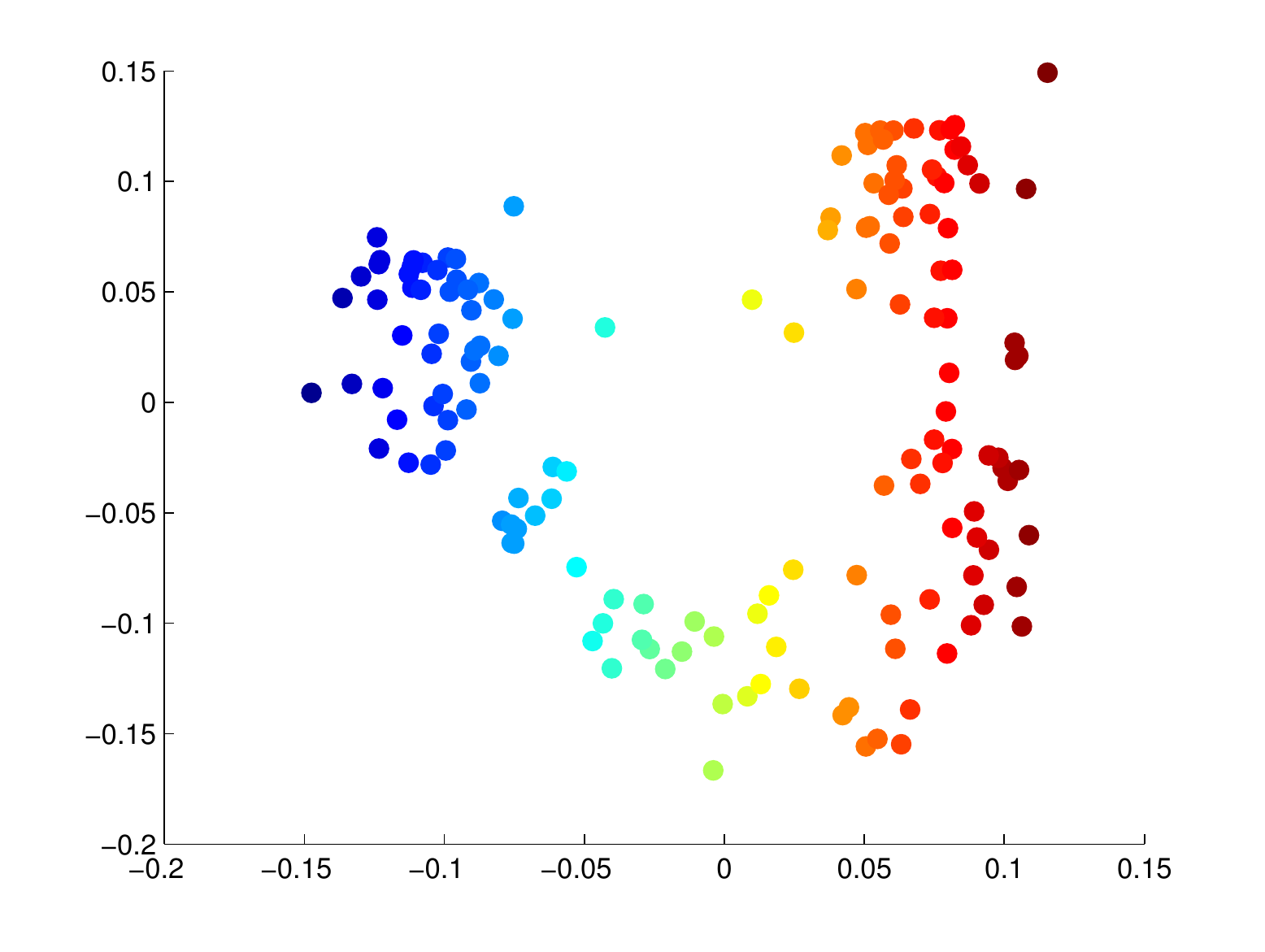}
\end{tabular}
	\caption{Embeddings conditioned on the absence of an edge at
          the center location. Note how less structured it is,
          compared to the positive embeddings. As such it could serve
          as a model for inhibitory connections, which span many orientations.}
\label{fig:embeddings-inhib}
\end{figure}

Our approach also has implications beyond excitatory connections for
boundary facilitation. 
We repeated our conditional spectral embedding, but now conditioned on the 
{\em absence} of an edge at the center
(Fig.~\ref{fig:embeddings-inhib}). This could provide a model for
inhibition, as clusters of edges in this embedding are likely to
co-occur conditioned on the absence of an edge at the center.  We find
that the embedding is only one dimensional, with the significant
eigenvector a monotonic function of orientation difference from the
center orientation.  Compared to excitatory connections, this suggests
that inhibition is relatively unstructured, and agrees with many
neurobiological studies.

\begin{figure}[h]
	\centering
	\begin{tabular}{cc}
	\includegraphics[width=.35\textwidth]{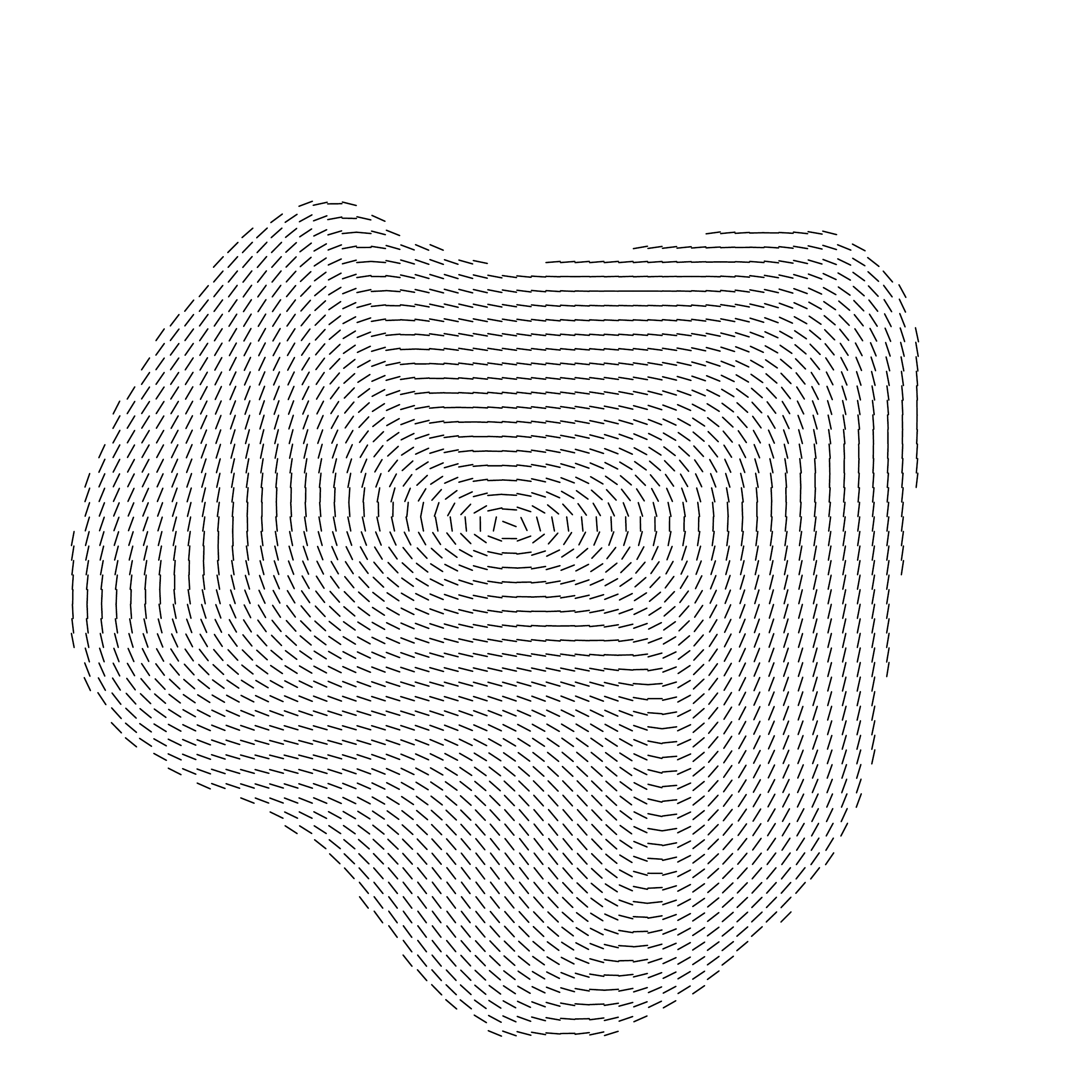} &
	\includegraphics[width=.35\textwidth]{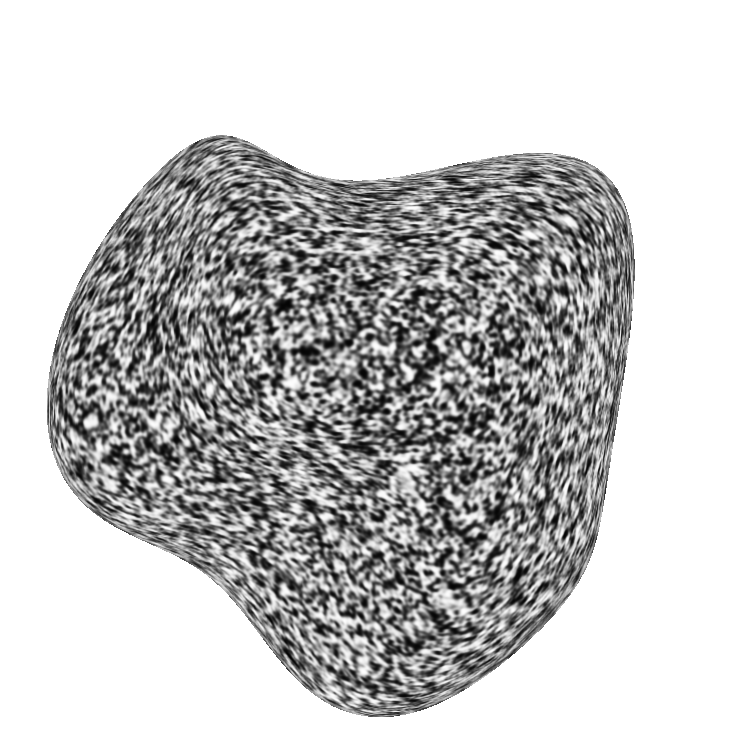} \\
  (a) & (b) \\
	\end{tabular}
	\begin{tabular}{ccc}
        \includegraphics[width=.2\textwidth,clip=true, trim=0 1.5 0 1.5]{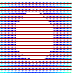} &
        \includegraphics[width=.2\textwidth,clip=true, trim=0 1.5 0 1.5]{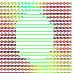} &
        \includegraphics[width=.2\textwidth,clip=true, trim=0 1.5 0 1.5]{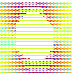} \\
         \includegraphics[width=.3\textwidth,clip=true, trim=0 1.5 0 1.5]{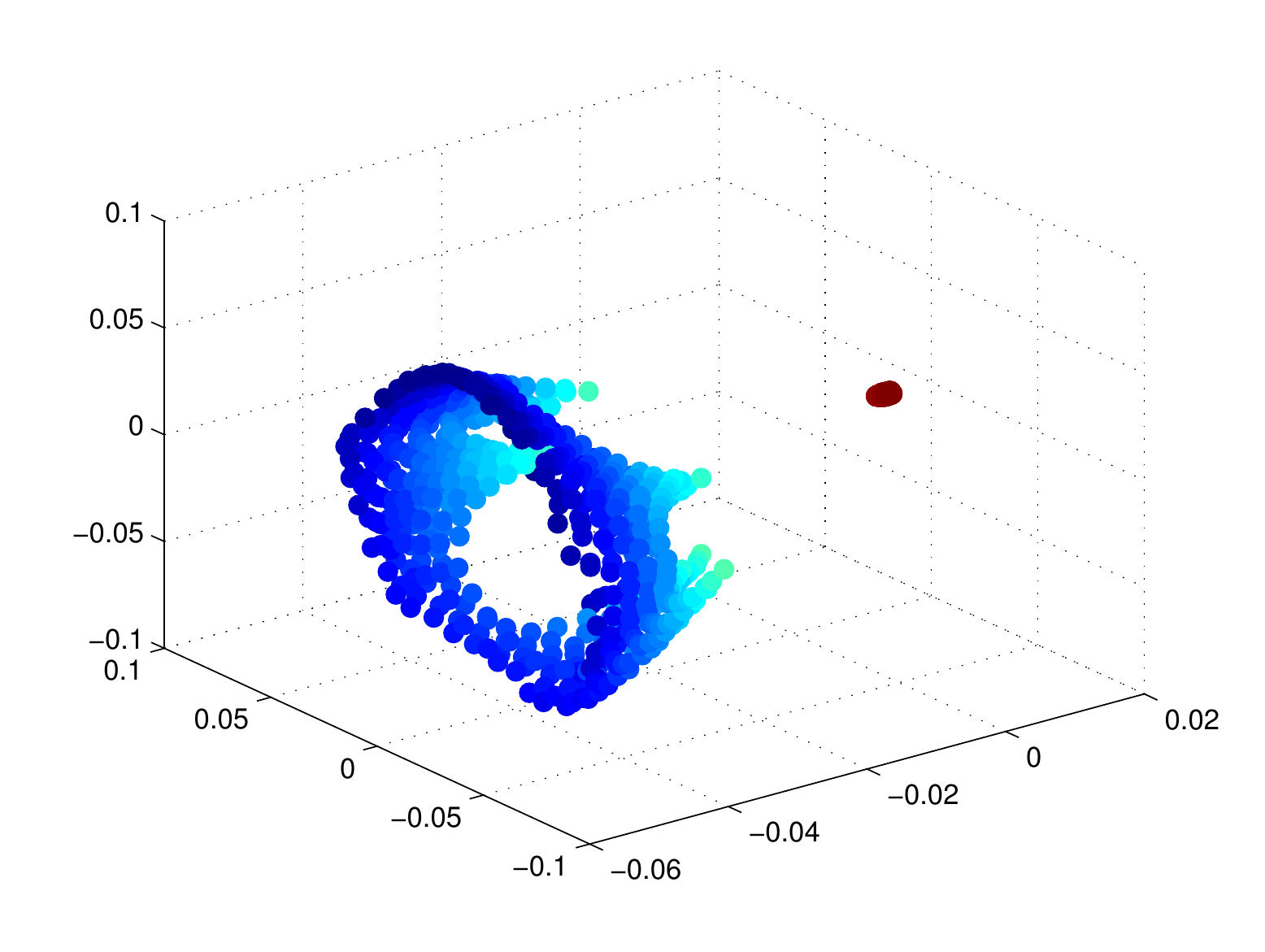} &
        \includegraphics[width=.3\textwidth,clip=true, trim=0 1.5 0 1.5]{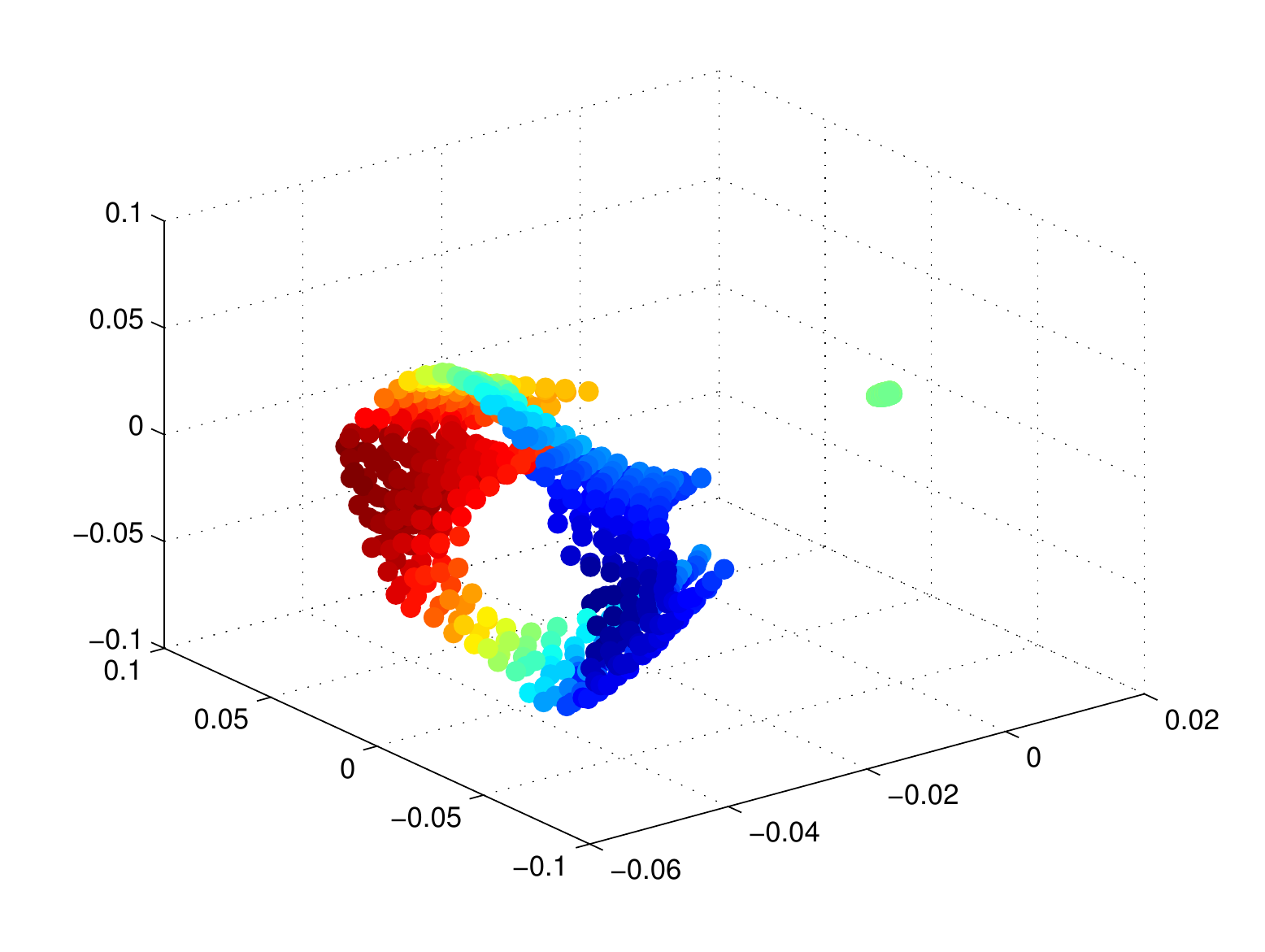} &
        \includegraphics[width=.3\textwidth,clip=true, trim=0 1.5 0 1.5]{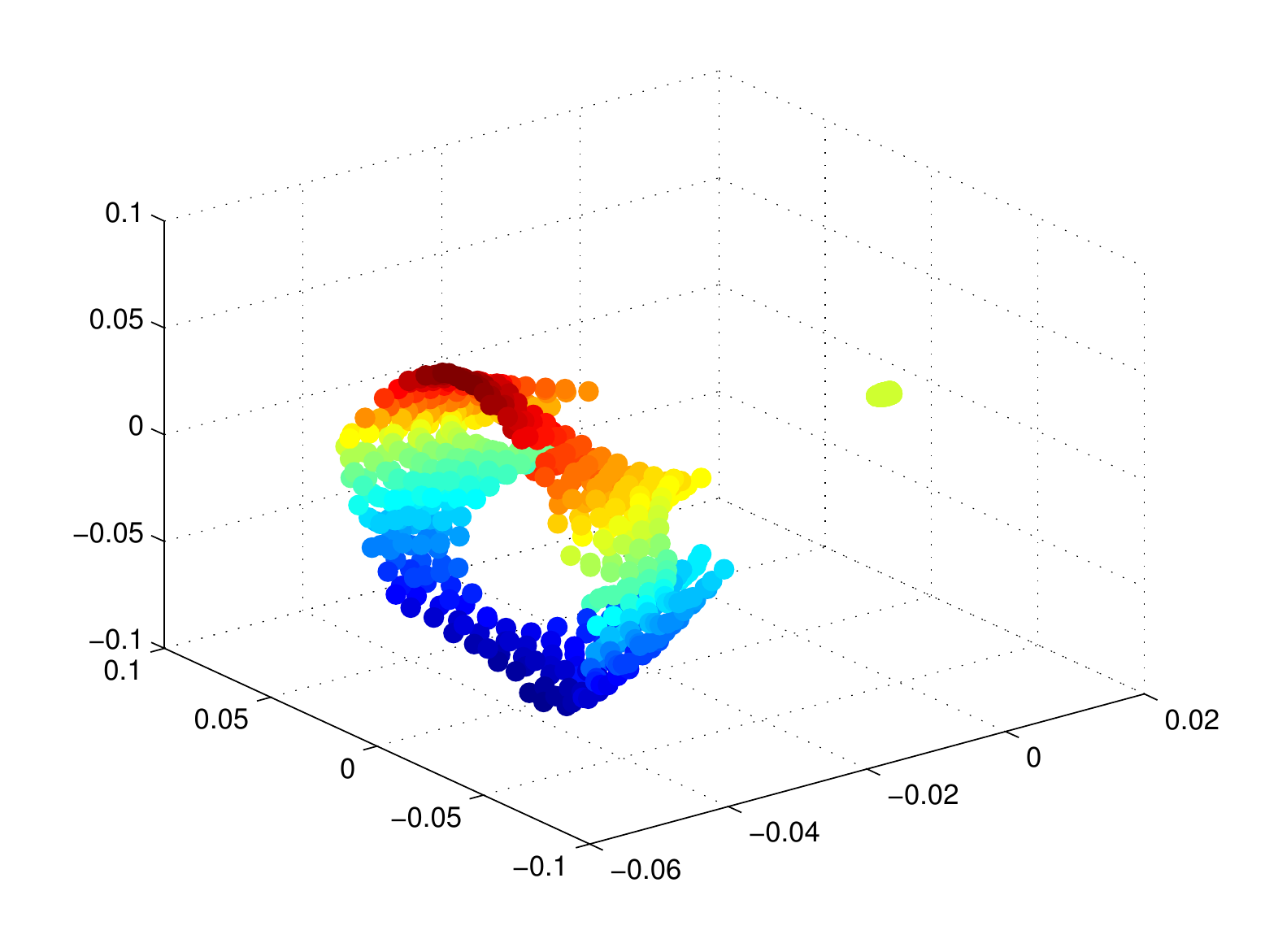}\\
        $\phi_2$ & $\phi_3$ & $\phi_4$
	\end{tabular}
	\caption{(top) Oriented textures provide information about
          surface shape. (bottom) 
As before, we looked at the conditional co-occurrence matrices of edge orientations over a series of randomly generated shapes.
Slant orientations and embedding colored by each eigenvector.  The edge map is 
	thresholded to contain only orientations of high probability.  The resulting embedding $\phi(v_i)$ of those orientations is shown below.  The eigenvectors of $P(i,j,0)$ are used to color both 
	the orientations and the embedding.  Clusters of orientations in this embedding have a \emph{high probability of co-occurring} along with the edge in the center.}
\label{fig:embeddings-shape}
\end{figure}

Finally, we repeated this third-order analysis (but  without local
non-maxima supression) on a structured model for isotropic textures on 
3D surfaces and again found a curvature dependency (Fig.~\ref{fig:embeddings-shape}). 
Every 3-D surface has a pair of associated dense texture flows in the image plane that correspond to the slant and tilt directions of the surface.  For isotropic textures, 
the slant direction corresponds to the most likely orientation
signalled by oriented filters.  Hence, the 
distribution of these directions in 3-D shape are an excellent proxy for measuring the response
of orientation filters to dense textured objects.

\begin{figure}[h]
	\begin{tabular}{cccc}
		\includegraphics[width=.25\textwidth]{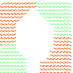} & 
		\includegraphics[width=.25\textwidth]{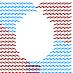} & 
		\includegraphics[width=.25\textwidth]{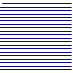} & 
		\includegraphics[width=.25\textwidth]{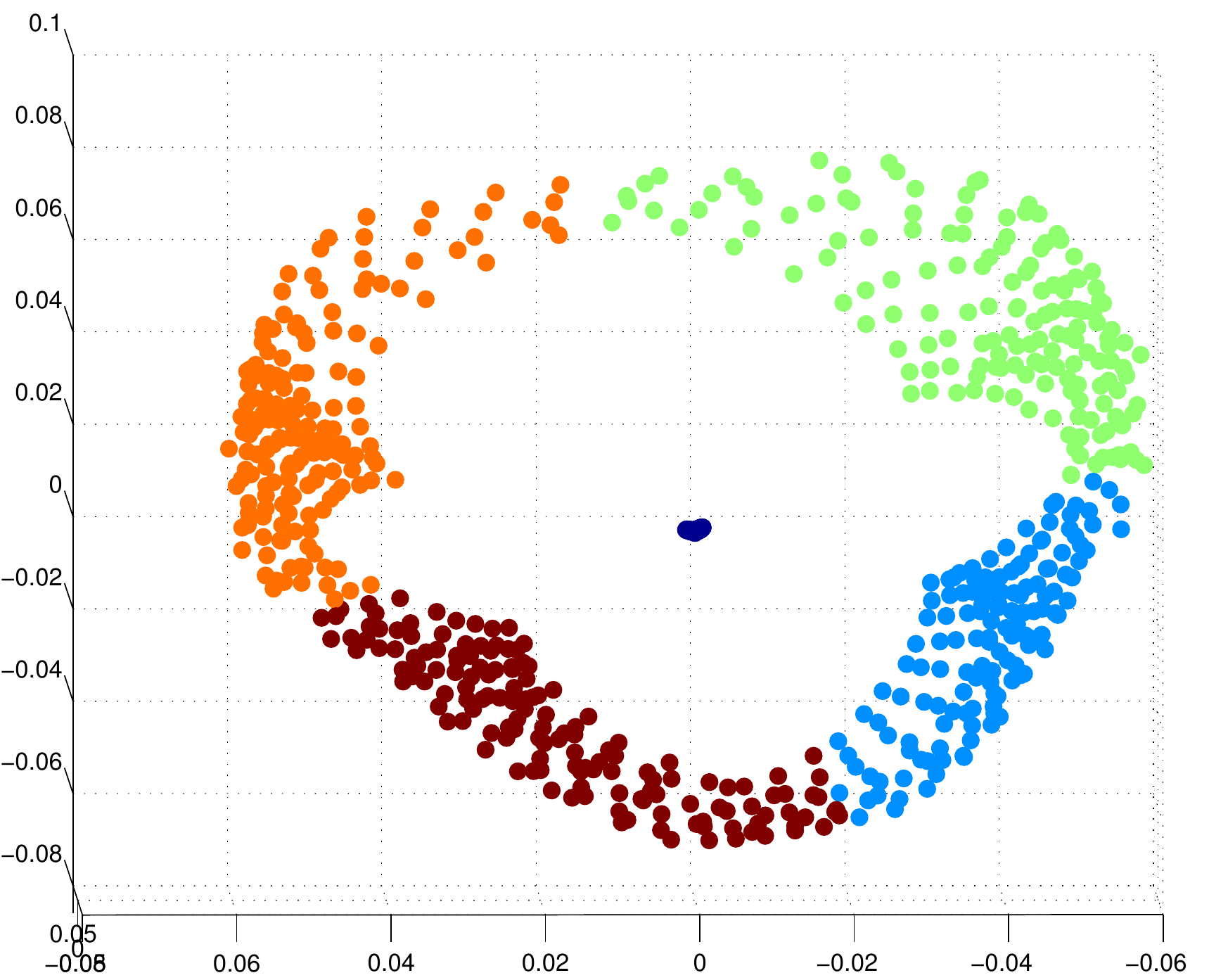}
	\end{tabular}
	\caption{Clustering of dense texture flows.  Color corresponds to the cluster index.  
	Clusters were separated into different figures so as to minimize the $x,y$ overlap 
	of the orientations.  Embedding on the right is identical to the embeddings above, but 
	viewed along the $\phi_3, \phi_4$ axes.}
\label{fig:embeddings-shape-cluster}
\end{figure}

As this is a representation of a dense vector field, it is more difficult to interpret than the
edge map.  We therefore applied k-means clustering in the embedded space and segmented the 
resulting vector field.  The resulting clusters show two-sided continuation of the texture flow
with a fixed tangential curvature (Fig.~\ref{fig:embeddings-shape-cluster}).

In summary, then, we have developed a method for revealing third-order
orientation structure by spectral methods. It is based on a diffusion
metric that makes third-order terms explicit, and yields a Euclidean distance
measure by which edges can be clustered. Given that long-range
horizontal connections are consistent with these clusters, how
biological learning algorithms converge to them remains an open question.
Given that research in computational neuroscience is turning to
third-order  \cite{ohiorhenuan2011}  and specialized interactions,
this question now becomes more pressing.
\vspace{60ex}






 \bibliographystyle{plain}
 \bibliography{Nipsbib}

\end{document}